\documentclass[aip,jap,reprint]{revtex4-1} 

\usepackage{amsmath}
\usepackage[normalem]{ulem}
\usepackage{amssymb}
\usepackage{latexsym}
\usepackage{indentfirst}
\usepackage{graphicx}
\usepackage{color}
\usepackage{bm}
\usepackage{multirow}

\begin{document}

\title{Bimodal switching field distributions in all-perpendicular spin-valve nanopillars}

\author{D.~B. Gopman}
\affiliation{Department of Physics, New York University, New
             York, NY 10003, USA}
\author{D. Bedau}
\affiliation{HGST San Jose Research Center,
             San Jose, CA 95135 USA}
\author{S. Mangin}
\affiliation{Institut Jean Lamour, UMR CNRS 7198 –Universit\'{e} de Lorraine, Nancy, France}
\author{E.~E. Fullerton}
\affiliation{CMRR, University of California at San Diego,
             La Jolla, CA 92093, USA}
\author{J.~A. Katine}
\affiliation{HGST San Jose Research Center,
             San Jose, CA 95135 USA}
\author{A.~D. Kent}
\affiliation{Department of Physics, New York University, New
             York, NY 10003, USA}
\begin{abstract}
Switching field measurements of the free layer element of 75~nm diameter spin-valve nanopillars reveal a bimodal distribution of switching fields at low temperatures (below 100~K). This result is inconsistent with a model of thermal activation over a single perpendicular anisotropy barrier. The correlation between antiparallel to parallel and parallel to antiparallel switching fields increases to nearly 50~\% at low temperatures. This reflects random fluctuation of the shift of the free layer hysteresis loop between two different magnitudes, which may originate from changes in the dipole field from the polarizing layer. The magnitude of the loop shift changes by 25\% and is correlated to transitions of the spin-valve into an antiparallel configuration.
\end{abstract}


\maketitle

\begin{figure}[b!]
\begin{minipage}{\textwidth}
\centering
    \includegraphics[width=6.in,
    keepaspectratio=True]
   {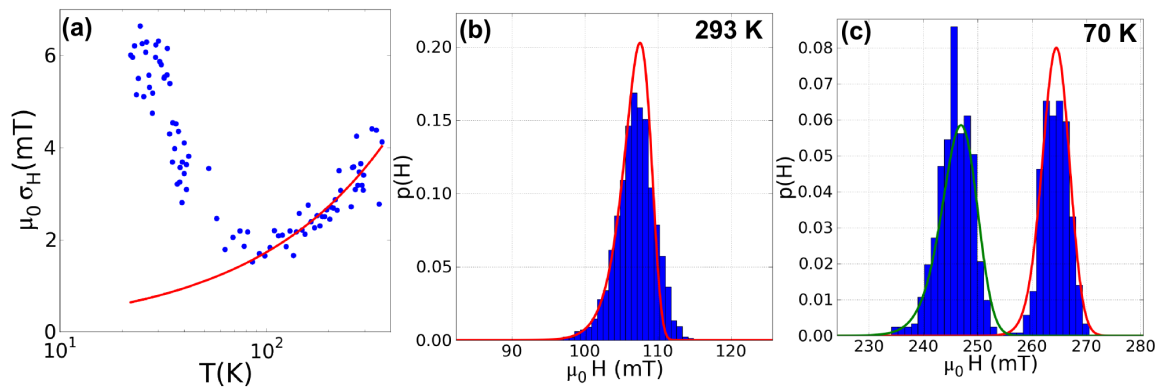}
\caption{\label{fig:Variance} (a) Temperature dependence of the switching field variance. The blue points are the experimental data and the red curve indicates the best-fit curve from the N\'{e}el-Brown thermal activation (TA) model. (b) Switching field histogram used to obtain variance at T=293~K with best-fit curve from TA model. (c) Histogram at T=70~K showing bimodal switching distribution with best-fit curves from TA model.}  
\end{minipage}
\end{figure}

Magnetization reversal in nanopillar spin-valves with all-perpendicular magnetizations has a direct impact on magnetic information storage technologies, such as magnetic random access memories.\cite{Kent2010,Brataas2012} Composed of ultrathin magnetic multilayers with tunable perpendicular anisotropy, spin-valves with lateral sizes down to tens of nanometers are being produced that are thermally stable at room temperature, with low critical switching currents.\cite{Mangin2006,Mangin2009,Ikeda2010} This geometry also gives rise to an out-of-plane dipole field from the polarizer, which can shift the center of the free layer minor hysteresis loop by a considerable fraction of the room temperature coercive field and has been shown to cause asymmetric reversal behavior for antiparallel to parallel ($AP \rightarrow P$) and $P \rightarrow AP$ transitions.\cite{Bernstein2011,Gopman2012}


We have recently investigated the thermally activated reversal of 75~nm spin-valve nanopillars to probe the barrier height to magnetization reversal.\cite{Gopman2013} Here we report measurements of the distribution of switching fields by conducting over 1,000 free layer hysteresis loops under a linearly swept magnetic field as a function of temperature. The switching field distributions at low temperatures (below 100~K) reveal the onset of a bimodal switching field distribution (compare Figs.~\ref{fig:Variance}(b)\&(c)). The bimodal switching distributions lead to the marked increase in switching field variance at low temperatures, as shown in Fig.~\ref{fig:Variance}(a). This behavior is inconsistent with a single energy barrier process described within the N\'{e}el-Brown model of magnetization reversal.\cite{Neel1949,Brown1963,Wernsdorfer1997,Sun2002}


In this paper we show that random fluctuations of the center of the free layer hysteresis loop are the source of the second mode. The coefficient of correlation between $AP \rightarrow P$ and $P \rightarrow AP$ switching fields increases with decreasing temperature, suggesting that changes in the loop shift become more significant at lower temperatures. Finally, we present further details at a representative temperature (70~K) in which we show the rate at which the hysteresis loop shift telegraphs between two values and indicate that changes in the shift occur more frequently following the $P \rightarrow AP$ transition. We conjecture that this could be due to changes in the magnetization of the second ferromagnetic layer (polarizer) induced by the free layer switching.



The 75~nm~diam nanopillars studied here are part of an all perpendicular spin-valve device consisting of a Co/Ni free layer and a Co/Ni and Co/Pt multilayered polarizer layer separated by a 4~nm Cu spacer. Details on materials and sample preparation have been reported previously.\cite{Gopman2013,Mangin2006} \clearpage Measurements were taken in a closed-cycle cryostat between the poles of an electromagnet oriented perpendicular to the device plane and at temperatures ranging from 20~K - 400~K. The reference layer magnetization switches for an applied field close to 1~T. Since no fields greater than 0.5~T are applied during the measurements, the reference layer is expected to remain stable.  

\begin{figure}[t!]
  \begin{center}
   \includegraphics[width=3.0in,
    keepaspectratio=True]
   {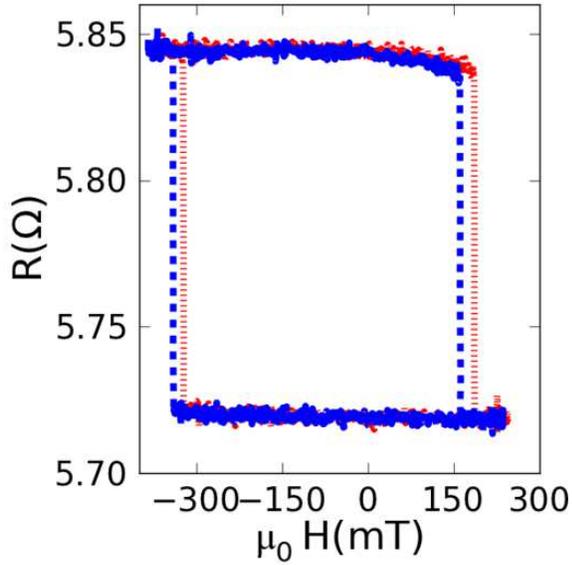}
  \end{center}
  \caption{\label{fig:Hysteresis} Sequential hysteresis loops of the free layer element of a 75~nm-diam nanopillar spin-valve at T=70~K. The first hysteresis loop (wide broken blue line) is more offset from zero applied field than the subsequent loop (narrow broken red line).}
\end{figure}

The magnetization of the free layer is probed indirectly with four-probe measurements of the differential resistance of the spin-valve device under an $50 \, \mu A$ excitation current using standard lock-in techniques.  Figure~\ref{fig:Hysteresis} shows two sequential resistance versus applied perpendicular field hysteresis loops at 70~K. The sharp changes in resistance indicates switching of the free layer into a parallel or antiparallel configuration with the reference layer. The approximately 85~mT shift of the center of the hysteresis loop denoted by the wider broken blue line drops to 65~mT in the immediately subsequent hysteresis loop denoted by the more narrow broken red line. These two distinct loop shifts are the source of the bimodal switching histogram in Fig.~\ref{fig:Variance}(c) and persist down to the lowest temperatures.

\begin{figure}[b!]
  \begin{center}
   \includegraphics[width=3.0in,
    keepaspectratio=True]
   {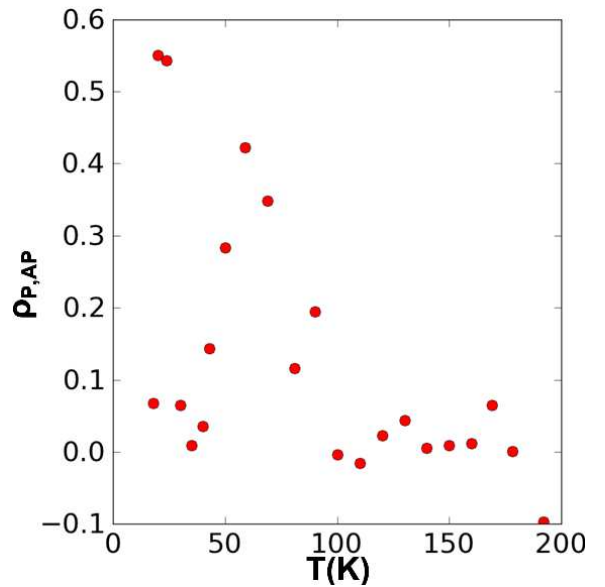}
  \end{center}
  \caption{\label{fig:Covariance} Correlation coefficient $\rho _{P, AP}$ for switching into the $P$ and $AP$ states as a function of temperature. The trend shows an increased correlation with decreasing temperature.}
\end{figure}

We present broader confirmation of this phenomenon by investigating the correlation between $AP \rightarrow P$ and $P \rightarrow AP$ switching fields with decreasing temperature. Figure~\ref{fig:Covariance} displays the correlation coefficient of the two switching branches. The correlation coefficient $\rho _{P, AP}$ is defined as:\begin{eqnarray}
&& \rho_ {P, AP} = \frac{ \sum _{i=1}^n (H^{P}_i - \overline{H}^{P})(H^{AP}_i - \overline{H}^{AP})}{(n-1) \sigma_{P} \sigma_{AP}},
\label{CorrelationCoeff}
\end{eqnarray}
where $H^{P(AP)}$ is the $AP \rightarrow P$ ($P \rightarrow AP$) switching field and $\sigma$ is the switching field variance for the transitions. When thermal activation dominates the switching process, the two switching field branches $AP\rightarrow P$ and $P\rightarrow AP$ should be uncorrelated. Nevertheless, we note a trend in which the correlation approaches 50\% with lower temperatures. As we will see below, the loop shifts do not always happen on the same branch of each hysteresis cycle, which explains why the switching field correlation always falls below 100\%.

\begin{figure}[t!]
  \begin{center}
   \includegraphics[width=3.0in,
    keepaspectratio=True]
   {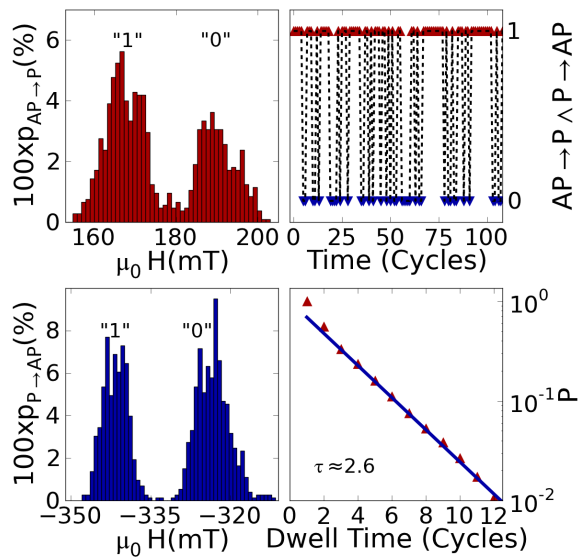}
  \end{center}
  \caption{\label{fig:LoopAnalysis} Switching field histograms at 70~K for (a)$AP \rightarrow P$ and (b) $P \rightarrow AP$ transitions. ``1''s (``0''s) reflect data belonging to the large (small) loop shift population.(c) Logical $AND$ ($AP \rightarrow P \wedge P \rightarrow AP$) for each field loop cycle indicates if the loop shift stays constant during an entire cycle. (d) Probability $P$ (red triangles) for a given shift (large or small) to endure versus duration. Exponential decay function (blue line) reflects a dwell time $ \tau \approx 2.6$ cycles.}
\end{figure}

We will now closely investigate the switching field data at a fixed temperature (70~K), which will provide a characteristic example of the changes in the loop shift across 1,000 hysteresis cycles. We can assign each switching event a switching index (``1'') or (``0'') according to one of the modes of the distribution in Figs.~\ref{fig:LoopAnalysis}(a)\&(b), which allows us to identify if a large or small loop shift preceded or coincided with each switching event. The cycles in which the loop shift remains ``large'' or ``small'' can be determined by taking the logical ``AND'' operation of the $AP \rightarrow P$ and $P \rightarrow AP$ switching index for each cycle (i.e. $1 \wedge 1 = 1$ \& $0 \wedge 0 = 1$ but $1 \wedge 0 = 0$ \& $0 \wedge 1 = 0$). Figure~\ref{fig:LoopAnalysis}(c) shows the effect of this logical operation - ``1''s for full hysteresis cycles with a fixed loop shift and ``0''s for cycles in which the shift changes between the $AP \rightarrow P$ and $P \rightarrow AP$ branches. We thus obtain the distribution of dwell times for the shifted and non-shifted states in Fig.~\ref{fig:LoopAnalysis}(d). Analogous to a telegraph signal in the time-domain, we fit the probability with an exponential decay law, from which we find the shift decay time $ \tau \approx 2.6$ cycles. We also investigate the conditional probabilities of a increase or decrease in the loop shift for $AP \rightarrow P$ versus $P \rightarrow AP$ transitions. The probability that the loop shift changes following the $P \rightarrow AP$ transition (0.436) is larger than following the $AP \rightarrow P$ transition (0.336). The conditional probability that the loop shift tends to decrease rather than increase after the $P \rightarrow AP$ transition (0.674) is markedly greater than for the $AP \rightarrow P$ transition (0.244).

The origin of the changes in the loop shift can not be immediately gleaned in these devices. For example, the resistance levels for $AP$ and $P$ spin-valve states do not change with the field shift as seen in Fig.~\ref{fig:Hysteresis}. We note from the conditional probabilities that the $P \rightarrow AP$ transition is typically preceding a reduction. We conjecture that this may be a dynamic effect associated with the free layer reversal, in which large local fields from a propagating domain wall in the free layer could create a small magnetic domain at an edge in the polarizer layer, similar to results seen in full-film spin-valves and spin-valve nanowires.\cite{Hauet2008,Mohseni2011,Liu2012} However, we cannot exclude alternative mechanisms for this fluctuating loop shift, including a small edge domain in the free layer or antiferromagnetic coupling within the free layer that might influence the free layer reversal behavior part of the time. 

The appearance of bimodal switching field distributions in nanopillar spin-valves reduces the reliability of device operation. For devices requiring efficient operation of spin-torque-induced switching, devices must be designed with narrow switching distributions. Changes in the loop shift adds noise to the system that can exceed the contribution of other random sources. Substantial changes in the magnitude of the loop shift may also reflect changes in the thermal stability of the free layer element in a way that could compromise device performance. 

\section*{Acknowledgments}
This research was supported at NYU by NSF Grant Nos. DMR-1006575 and NSF-DMR-1309202, as well as the Partner University Fund (PUF) of the Embassy of France. Research at UL supported by ANR-10-BLANC-1005 ``Friends'', the European Project (OP2M FP7-IOF-2011-298060) and the Region Lorraine. Work at UCSD supported by NSF Grant No. DMR-1008654. 

\bibliographystyle{apsrev}

\end{document}